\documentclass[eqsecnum,twocolumn,aps,a4paper]{revtex4}

\usepackage{graphicx}%
\usepackage{dcolumn}
\usepackage{amsmath}
\usepackage{latexsym}

\def\al{CBr$_{4}$-C$_{2}$Cl$_{6}$}
\def\cb{CBr$_{4}$}
\def\cc{C$_{2}$Cl$_{6}$}

\def\mis{~$\mu ms^{-1}$}

\def\degres{$^{\rm o}$}
\def\mic{$\rm{\mu{}m}$}  
\begin{document}

\title{An experimental method for the {\it{in-situ}} observation of eutectic growth patterns in bulk samples of transparent alloys}

\author{Sabine Bottin-Rousseau, Mika\"el Perrut, Christian Picard, Silv\`ere Akamatsu and Gabriel Faivre}

\affiliation{Institut des Nanosciences de Paris, CNRS UMR 7588,
Universit\'e  Pierre-et-Marie-Curie (Paris VI) and Universit\'e Denis-Diderot (Paris VII), Campus 
Boucicaut, 140 rue
de Lourmel, 75015 Paris, }
               
\email{sabine.bottin-rousseau@insp.jussieu.fr}
 
\date{\today}

\begin{abstract}
We present an experimental method for the  {\it{in-situ}} observation of directional-solidification fronts in bulk samples of transparent eutectic alloys.  The growth front is observed obliquely in dark field through the liquid and a glass wall of the container with a long-distance microscope. We show that a focused image of the whole growth front can be obtained at a certain tilt angle of the microscope.  At this tilt angle, eutectic fibers of  about 3.5\mic~in diameter can be clearly seen over the whole growth front in 400-\mic~thick samples.
\end{abstract}

\pacs{PACS numbers: }
 
\maketitle

\section{Introduction}

Directionally solidified eutectic alloys form extended growth patterns of spacing in the 1-10\mic~range for solidification rates on the order of 0.1\mis.  The dynamics of these patterns has been a question of fundamental and practical interest for decades. Progress in this field requires experimental methods permitting a spatiotemporal follow-up of growth patterns during solidification.  Transparent nonfaceted eutectic alloys, such as \al~or Succinonitrile-(d)Camphor (SCN-DC), have been introduced for this purpose a long time ago \cite{HuntJacks66}. Usually, these alloys are  directionally solidified in very thin (10-\mic~thick, typically) samples. This confers a strongly one-dimensional (1D) character  to the growth patterns, which can then  be observed in side view, {\it{i.e.}} along the direction normal to the sample plane, with a standard optical microscope. This thin-sample method has proved quite accurate  \cite{Ginibre97, akaplapp02, scndc}, but can naturally be used only in the study of 1D eutectic growth patterns.  A method of observation  {\it{in situ}} applicable to 2D eutectic growth patterns has long been lacking. Recently, we published observations obtained with a new  method permitting  a spatiotemporal follow-up of growth patterns in  transparent samples of thicknesses ranging from 100 to 500\mic~  \cite{zigzag,philmag}. In this article, we present this experimental method in detail.

\begin{figure}[ht]
\includegraphics[width=5.2cm]{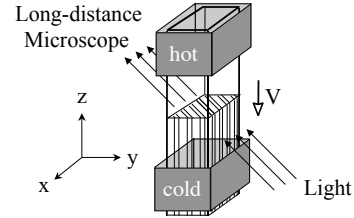}
\caption{Oblique observation of a transparent eutectic directional-solidification front. $z$: direction of the thermal gradient. The sample is pulled in the direction $-z$ at a rate $V$. $y$: normal to the sample walls. The optic axis of the microscope lies in the plane $yz$.  }
\label{sketch}
\end{figure}

\begin{figure}[htbp]
\includegraphics[width=7.5cm]{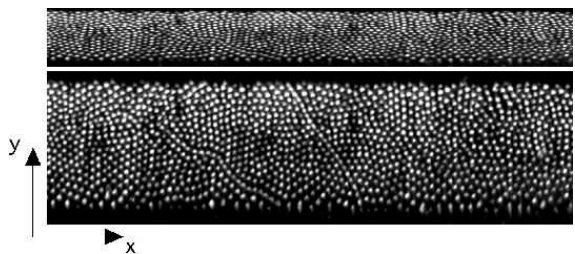} 
\caption{ Upper panel: Oblique view of a fibrous eutectic pattern in a 340-$\mu \rm{m}$ thick sample of eutectic SCN-DC. $V= 0.035 \mu \rm{ms^{-1}}$. Bright spots correspond to DC fibers, while the SCN matrix appears dark. The tilt angle of the LD microscope is about 40\degres. Lower panel:  rescaled image.    Horizontal dimension: $1300 \mu \rm{m}$.}
\label{rods}
\end{figure}

Figure  \ref{sketch} shows  the principle of the method.  A  directional-solidification sample of rectangular cross-section receives light through a wall of the crucible, and is observed with a tilted long-distance (LD) microscope through the opposite wall. The incident light beam and the LD microscope are oriented in such a way that the microscope collects only the light rays that emerge from  one of the solid phases, of which the growth front is composed. A dark-field image of, essentially, the sole growth front is thus obtained.  An example is  given in Figure \ref{rods}, which shows a  fibrous eutectic  growth pattern in a 340-\mic~thick sample of SCN-DC. It can be seen that fibers about 10\mic~in diameter are well resolved over the whole growth front. The purpose of this article is to explain how such an image can be obtained in spite of the large tilt angle of the LD microscope with respect to the growth direction.

The article is organized as follows. A technical description of the experimental setup is given in Section \ref{Technical}. Some data about the transparent alloys used are reported in  Section \ref{alloys}. The basic imaging properties of the setup are explained in Section \ref{Imaging}. The nature  of dark-field images is discussed in Section \ref{darkfield}. Section \ref{thermalbias} is devoted to the important question of the elimination of thermal bias, {\it{i.e.}} the fine control of the orthogonality between the isotherms and the growth axis.

\section{Technical description of the experimental setup} \label{Technical}

\begin{figure}[htbp]
\includegraphics[width=7.5cm]{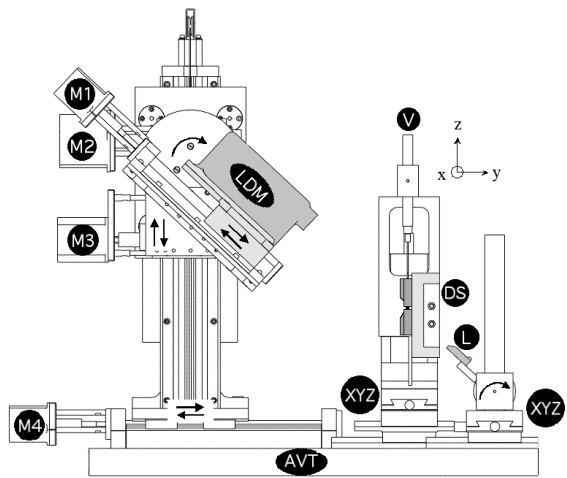}
 \caption{Technical sketch of the experimental setup.   DS: directional-solidification bench (see Fig. \ref{fours}).  $V$: dc motor. XYZ: motorized $xyz$ translation stages.   LDM: long-distance microscope (the video camera is not shown).  M1 to M4: step motors.  M1: control of the working distance.  M2: rotation of the microscope.  M3 and M4: automatic adjustment of focus.  AVT: anti-vibrating table. L:  linear fiberoptic array (the halogen lamp is not shown). The  reference frame is the same as  in Fig.  \ref{sketch}}
\label{seba}
\end{figure}

The experimental setup is composed of three distinct parts (Fig. \ref{seba}): a  directional-solidification (DS) bench, a light source, and a LD microscope (Questar QM100)  equipped with a numerical black-and-white video camera (Scion Corporation). The images are stored in a personal computer for further numerical treatment (contrast enhancement, and "rescaling", {\it{i.e.}} stretching of the captured image in the direction $y$ so as to restore its width  to the actual value $w$; see Fig. \ref{rods}). The DS bench is put vertically, with the direction  $z$  pointing upwards. The LD microscope and the light source are mounted on rotation stages of axes  parallel to $x$.  The different parts of the setup are mounted on independent translation stages, which are fixed on an  anti-vibrating table.  

\subsection{Directional-solidification bench} \label{bench}

\begin{figure}[ht]
\includegraphics[width=8cm]{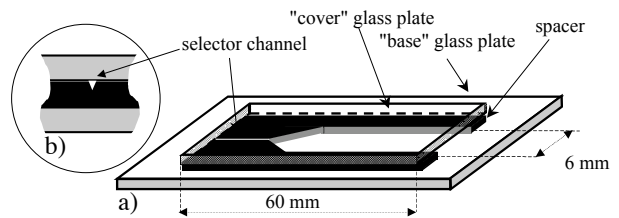}
\caption{a) Sketch of a crucible (glue seals are not shown).  b) Cross-section of the grain selector. }
\label{Echantillon}
\end{figure}

The DS bench has been designed for low melting point transparent alloys. In this study, both the \al~and SCN-DC eutectics were utilized.   The crucibles are made of  two 300-$\mu m$ thick glass plates (a  "base" and a "cover" plate) separated by plastic (PET) spacers (Fig. \ref{Echantillon}). The spacers are cut to a special funnel shape  in order to form a grain selector. The cover plate and the spacers are fixed to the base plate with a UV polymerisable glue (Norland NOA81). A  pressure is applied onto the cover plate during  the gluing process. The thickness $w$ of the inner volume depends on both the spacer thickness and the applied pressure.  We determine $w$ by measuring the outer thickness of the sample at the end of the process.  The crucibles are heated above the liquidus temperature of the alloy, filled with liquid alloy by capillary action under a controlled Ar atmosphere ($\approx 10\, \rm{mbar}$), and sealed after a rapid cooling to room temperature. The alloy is then entirely solid. The base glass plate fits in a rigid holder, which is inserted into the DS bench and connected to the translation motor.

\begin{figure}[htbp]
\includegraphics[width=7.5cm]{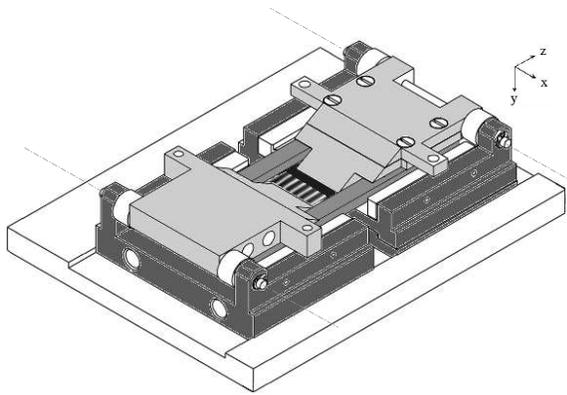}
 \caption{Directional-solidification bench. Each oven is made of a base (dark grey) and a cover (soft grey) copper block. A sample fixed in its holder (mid grey) is represented. The connection to the motor is not shown. Dash-dotted lines: rotation axes of the cover blocks.}
\label{fours}
\end{figure}

The DS bench is composed of two ovens  fixed on an insulating soil, and a translation motor. Each oven is made of a fixed (base),  and a removable (cover) copper block, between which the sample is inserted  (Fig. \ref{fours}). The blocks  are machined with V-shaped edges to permit oblique observation and illumination through the gap separating the ovens.  The base glass plate of the crucible is in contact with the base blocks of the ovens. A pressure is exerted onto the cover blocks by means of a screw-and-spring system (not shown) in order to ensure a better thermal contact and an more accurate guiding of the  sample translation. The temperature field is established by heat diffusion along the sample. The temperature gradient $G$  at the growth front can be varied from 40 to 110  $\pm 10 \%~\rm{Kcm^{-1}}$ by varying the width of the observation window. Temperature regulation is ensured by  water circulation for the cold oven, and by resistive sheets connected to an electronic PID controller for the hot oven. The temperatures  of the base and cover hot block are regulated independently of each other to permit the adjustment of the $y$ component of the thermal gradient (see Section \ref{thermalbias}). A dc motor (Physik Instrumente) ensures the translation of the sample at  velocities ranging from 0.01 to 10 $\mu \rm{m.s^{-1}}$ with an accuracy of $\pm 1 \%$. 

\subsection{Optical system}

\begin{figure*}[htbp]
\centerline{\includegraphics[width=17cm]{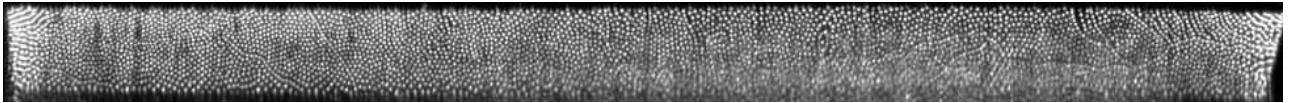} }
\caption{Whole-front mosaic micrograph of the same growth front as in Fig. \ref{rods}. Note the  lamellar growth patterns near the edges of the sample. Horizontal dimension: $4800 \mu \rm{m}$.}
\label{panoentier}
\end{figure*}

The Questar QM100 system is a Maksutov-Cassegrain catadioptric telescope utilizable at working distances ranging from 15 to 35~cm; at the working distance of 15~cm, the numerical aperture of the microscope is of 0.14, the diffraction-limited  resolution is  of 1.1\mic~and, and the depth of field is of 28\mic  (information from the constructor).  We usually set the working distance at 15~cm. The field of view of the complete instrument (LD microscope plus video camera) is 440-$\mu \rm{m}$  wide in the direction $x$. The whole growth front can be scanned along $x$ using the motorized translation stage of the DS bench  (Fig. \ref{panoentier}).

The tilt angle $\theta_t$ of the microscope axis, called "observation angle", is measured from the horizontal (side-view) position. It can be varied continuously from  0~to 60\degres. An automatized mechanical device comprising  two translation motors keeps the field of view and the focus unchanged as $\theta_t$ varies.  An additional translation stage serves to vary the working distance, {\it{i.e.}} to zoom or unzoom the image. An iris diaphragm is situated in a plane close to the entry of the LD microscope. Incidentally, a slight misalignment of the microscope axis with respect to the plane $yz$ forced us to decenter  this diaphragm in the direction $x$ in order to obtain a satisfactory dark-field contrast.

The light source is a $30\times2 \rm{mm}$ linear fiberoptic array  aligned on the direction $x$ illuminated by a halogen lamp. It can be rotated around $x$, and displaced in the  directions $y$ and $z$. For a horizontal distance between the source and the object of 5~cm, the angular aperture of the incidence beam is of about 2\degres~ in the plane $yz$, and 30\degres~ in the direction $x$ .  We adjust the incidence angle  $\theta_o$  by changing the vertical position of the source. Simultaneously, we rotate the source in order to keep the intensity of the incident light at its maximum.

\section{Transparent nonfaceted eutectic alloys} \label{alloys}

The two eutectic solid phases of the  \al~system, denoted $\alpha$ and $\beta$, are solid solutions of fcc and bcc structures, respectively. The two eutectic solid phases of the  SCN-DC system are practically pure bcc SCN , and hexagonal DC, respectively, and will therefore be denoted "SCN" and "DC".   The concentrations and temperatures at the  eutectic points in these two systems are given in Table \ref{indextable}. More details about the phase diagrams can be found elsewhere \cite{MerFai93, WitusiewiczActa04}. Here we are interested in the refractive indices of the eutectic phases, especially, the differences in refractive index between eutectic solids and liquid, in these systems. These refractive indices have never been measured, to our knowledge, so we attempted to calculate them using the Lorentz-Lorenz formula \cite{Hartsh64}. The ingredients of the calculation are the molecular refractivity of the components, and the density of the phases of interest.  We calculated the  molecular refractivities of the components from the measured values of the refractive indices of pure \cb, \cc, SCN and DC  available in the literature  \cite{Hanbook}. The densities  of the pure substances  at the temperatures of interest could also be calculated, or estimated with good accuracy from data available in the literature. Fortunately, all the phases of interest except one (the $\beta$ phase of \al) have a low concentration in one of the components. We estimated their density (and hence their refractive index) assuming that the molar volumes in these phases were the same as in the pure substances (Table \ref{indextable}). This assumption is correct in the case of the SCN and DC phases, reasonable in the case of the two liquids and the $\alpha$ phase, but questionable in the case of the $\beta$ phase. The calculated difference in refractive index is -.005 for the SCN-liquid pair, 0.087 for the DC-liquid pair. This strong difference in order of magnitude is supported by experimental observations, which show that, in thin samples, the contrast of the DC-liquid interfaces much stronger than that of the SCN-liquid interfaces  \cite{ scndc}. Similarly, experimental observations in thin samples of \al~show a strong contrast of the  $\alpha$-liquid interfaces and a weak contrast of the $\beta$-liquid interfaces  \cite{akaplapp02}, indicating that the  refractive index of $\beta$ in Table \ref{indextable} is most probably largely overestimated.

\begin{table*}[htbp]
\begin{tabular}{|c|c|c|c|c|c|c|c|}
\hline
 Material & $T_E$ & Concentration  & density  & Refractive & $\theta_{\perp}$ &  $f_{\perp}$&  $R_{\perp}$ \\
 & (C) &($mol \%$) & ($g/cm^3$) &index & ($degree$) &&  \\
\hline
SCN-DC  & 38.4 & & &   &    49.8        &  0.40     &   0.60   \\
SCN  & &     0       &  1.046   &     1.446  & & &\\
Liquid &  &  12.4     &  1,002 &      1.451       &           &      &     \\
DC &  &    100      &  0.977   &   1.538  & & & \\
\hline
\al  & 84.4 & & &   &  48.6        &   0.35  &   0.53 \\
$\alpha$ &&    8.5    &  3.03  &  1.622  & && \\
Liquid & &  11.8     &  2.85 &          1.582     &     &    &     \\
$\beta$ & &    19      &    (2.92)&      (1.61)& && \\
\hline
\end{tabular}
\caption{Refractive index  of the equilibrium eutectic phases in the alloys SCN-DC and  \al~calculated using the Lorenz-Lorentz formula.  $T_E$: eutectic temperature. $\alpha$, $\beta$: solid phases in equilibrium with the liquid at $T_E$.  $\theta_{\perp}$,  $f_{\perp}$,  $R_{\perp}$: see Section \ref{Imaging}.}
\label{indextable}
\end{table*}

\section{Imaging properties} \label{Imaging}

\subsection{Optimal observation angle}

\begin{figure}[htbp]
\includegraphics[width = 8cm]{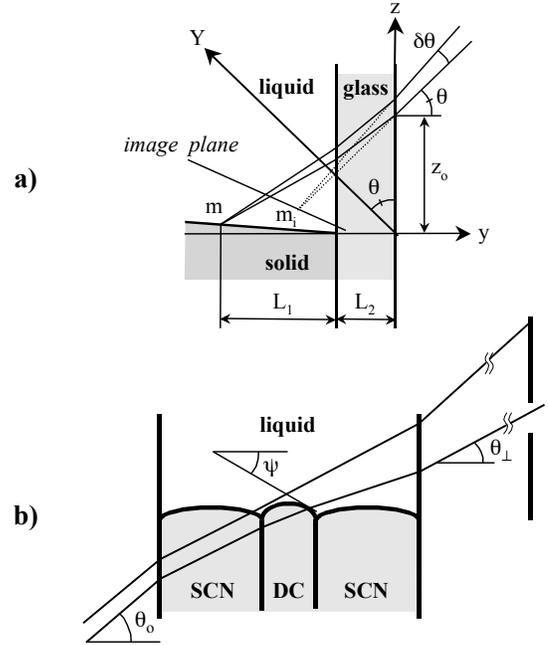}
 \caption{Sketches illustrating the processes under study. a) Large-scale image of a planar growth front. (The direction of $y$  is opposite to that defined in Fig. \ref{sketch}). b) Origin of dark-field contrast in the case of a fibrous SCN-DC pattern.} 
 \label{schemaImage}    
\end{figure}

Two basic properties of the experimental setup are the existence of an optimal observation angle, at which an in-focus image of the whole growth front can be obtained, and a relatively small astigmatism aberration. To explain these properties, it will suffice to consider the propagation of the light rays in the plane $yz$ (Figure \ref{schemaImage}a).  We consider a rectilinear object (a planar growth front) immersed in a  liquid of refractive index $n_1$ separated from the external medium by a glass plate of refractive index  $n_2$ and thickness  $L_2$. We denote by 2$\epsilon$ the angular aperture of the entry diaphragm. We consider a light ray emitted by a point $m$ of the object, and emerging in the external medium with a slope  $p=tan\theta$. The slopes $p_1=tan\theta_1$  and $p_2=tan\theta_2$ of the light ray in the liquid and the glass, respectively, are related to $p$ by Snell's law
 
 \begin{equation}
 \label{snell}
n_1  \sin \theta_1 =n_2  \sin \theta_2 = \sin \theta .
 \end{equation}
The equation for the light ray in the external medium reads
 \begin{equation}
 \label{ray1}
z =z_o +  py ,
 \end{equation}
where
 \begin{equation}
\label{ray2}
z_o=L_1(p_1+ p_{b}) +L_2p_2 ,
 \end{equation}
$L_1$ is the distance  of  $m$ from the cover plate, and   $p_{b}$  is the slope of the growth front  due to thermal bias. The  intersection of the light rays of slopes $p$ and  $p+\delta p$ emitted by $m$ is given by
 \begin{equation}
\label{intersec1}
z =z_o(L_1,p) +  py=z_o(L_1,p+\delta p) + (p+\delta p)y .
 \end{equation}
The image $m_i=(y_i,z_i  )$ of $m$ corresponds to the limit $\delta p\rightarrow0$. One has
 \begin{equation}
\label{yi}
y_i =-\partial z_o /\partial p\mid_{p=p_t},
 \end{equation}
and $z_i =z_o +  p_t y_i$,  where  $p_t=tan\theta_t$.  The image of the growth front is a curve of slope
\begin{equation}
\label{pi}
p_i =\partial z_i /\partial  y_i\mid_{p=p_t}=p_t-\frac{p_{1t}+ p_{b}}{\partial p_{1} /\partial p_t} .
 \end{equation}
Since   $p_i$ is independent of $L_1$, the image of a planar front is a plane. 

A  sharply focused image of the whole front can be obtained only when the image plane is normal to the axis of the microscope, {\it{i.e.}} when $p_ip_t=-1$. This condition defines the optimal observation angle $\theta_{\perp}$. By setting $p_i=-p_t^{-1}$ in eq. \ref{pi}, we obtain the following equation for $\theta_{\perp}$: 
 \begin{equation}
\label{sf}
p(p_1+ p_{b})=(1+p^2)\partial p_1 /\partial p .
 \end{equation}
For an arbitrary value of $\theta$, the contraction factor of the image in the direction  $Y$ normal to $x$ and the microscope axis is 
\begin{equation}
\label{factor}
f=\partial z_o/ \partial L_1\cos \theta \cos \psi . 
 \end{equation}
From eq. \ref{ray2} and the definition of $p$, one obtains $f\propto(p_1+ p_{b})(1+p^2)^{-1/2}$, and hence $\partial f/ \partial p\propto\partial p_1 /\partial p (1+p^2)-(p_1+ p_{b})p$. Comparing this formula with eq. \ref{sf} shows that  $f$ is maximum at the observation angle $\theta_{\perp}$.

It is useful to note that, when the growth front is normal to $z$ ($ p_{b}=0$), eq. \ref{sf} has the simple form
 \begin{equation}
\label{tperp}
\tan \theta_{\perp}=(1-n_{1}^{-2})^{-1/4} .
 \end{equation}
The contraction factor for this value of $\theta$ is
\begin{equation}
\label{fperp}
f_{\perp}=n_1^{-1}\sin^2\theta_{\perp} . 
 \end{equation}
These functions vary slowly with $n_1$, except when $n_1$ is close to unity.  Their estimated values at the eutectic points of the  \al~and SCN-DC alloys are displayed in Table \ref{indextable}. 

\subsection{Astigmatism} \label{Astigm}

\begin{figure}[htbp]
\includegraphics[width=7cm]{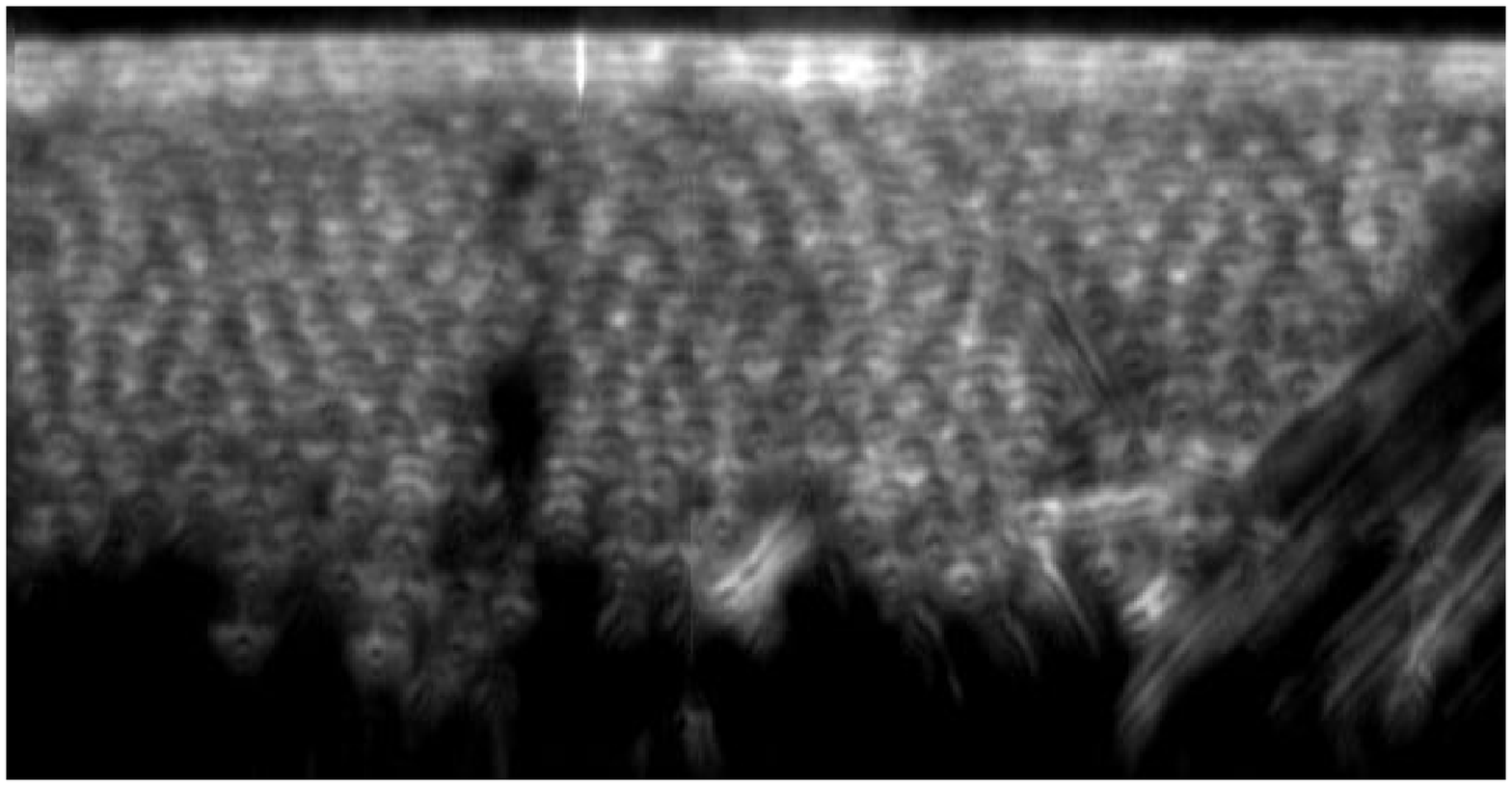}
 \includegraphics[width=7cm]{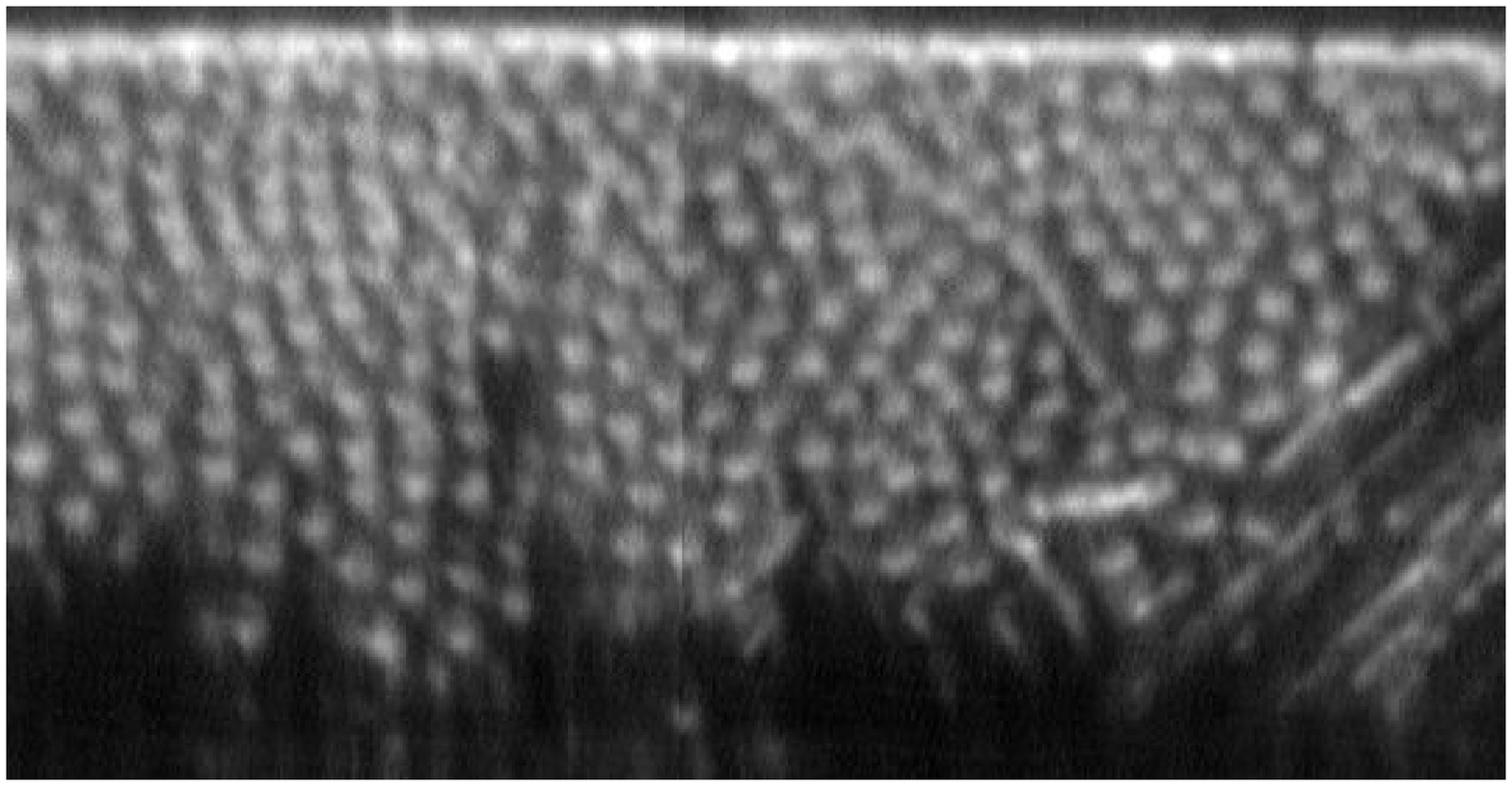}
\caption{A fibrous pattern in a 350-$\mu \rm{m}$ thick
\al~sample viewed without (upper panel)  and with (lower panel) an additional diaphragm. $V= 0.25 \mu \rm{ms^{-1}}$. Rescaled images. Horizontal dimension: $750 \mu \rm{m}$.} 
\label{Diaph}
\end{figure}

It is a well-known fact that the astigmatism aberration of a planar diopter  is an increasing function of the distance of the object from the diopter and the aperture of the optical setup. This is the reason for which we took a thin cover glass plate and introduced an additional entry diaphragm. Indeed, the  images obtained with the sole LD microscope ($\epsilon \approx 8 \rm{^o}$ at a working distance of 15~cm) had  a blurry aspect, which disappeared when $\epsilon$ was reduced by means of the entry diaphragm (Fig. \ref{Diaph}). Reasonably sharp images were obtained for $\epsilon=4.4\rm{^o}$ and $w\approx400\mu \rm{m}$ (see Fig. \ref{rods}) indicating that the astigmatism aberration of the setup was comparable to the diffraction-limited resolution of the LD at microscope  (Å2.2~\mic) at this value of $\epsilon$. 

We tested the validity of these empirical conclusions using eq. \ref{yi}. According to this equation, the shift from the sharp-focus point due to a deviation $\delta p$ from the chief ray is $\delta y_i=-1/2~\partial^{2} z_o /\partial p^{2}~\delta p$ along the axis $y$, and $\delta y_i\times\delta p/\cos\theta$ along the axis $Y$. Thus the broadening $\Delta Y$ of the image of $m$ due to astigmatism is
\begin{equation}
\label{deltaY}
\Delta Y\approx\frac{\mid\partial^{2} z_o /\partial p^{2}\mid}{2\cos^{3}\theta}\epsilon^2 .
 \end{equation}
Assuming that $n_2\approx n_1$, we obtain
\begin{equation}
\label{deltaYperp}
\Delta Y\approx R_{\perp}(L_1+L_2)\epsilon^2 .
 \end{equation}
where $R_{\perp}=1.5~n_{1}^{-1}\sin^2\theta_{\perp}$. For $n_1=1.451$, $L_1+L_2=750~\mu \rm{m}$, and  $\Delta Y=2.2~\mu \rm{m}$, we obtain  $\epsilon= 4.0 \rm{^o}$, in reasonable agreement with the above conclusions.

On the other hand, observations showed that, still for  $\epsilon=4.4\rm{^o}$ and $L_1<400\mu \rm{m}$, the image quality remained tolerable  for deviations of the observation angle from the optimum value as large as 10$\rm{^o}$. For this deviation and  $L_1=400\mu \rm{m}$, the maximum distance between a point of the image and $Y$ is of about 70\mic, which is comparable to the diffraction-limited depth of field of the LD microscope  (90\mic) at  the value of $\epsilon$  under consideration .

\section{Dark-field images} \label{darkfield}

\begin{figure}[htbp]
\includegraphics[width=7cm]{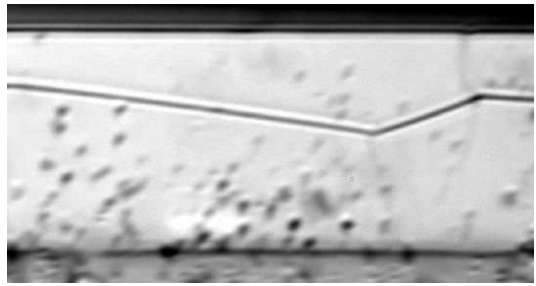}
\caption{Solidification front of a  $\alpha$ crystal  in a 400-$\mu \rm{m}$ thick sample of a dilute \al~alloy.   $V=2 $\mis. Non-rescaled image. A dark zigzag line runs along the intersection of a grainboundary with the solid-liquid interface. Angular points on this line are junctions with subboundaries. Dark spots are liquid droplets migrating  in the solid. The residual brightness appearing under the interface is due to multiple diffusion by these droplets. Horizontal dimension: 425 \mis. }
\label{dilue}
\end{figure}

When the solid and the liquid are homogeneous,  only the light rays that intersect the solid-liquid interface undergo a deviation. This deviation is a function of the difference in refractive index  between solid and liquid, and the slope angle $\psi$ of the interface. At  fixed observation angle,  a dark-field image is obtained by adjusting the  incidence angle  so as to capture only the light rays that suffer a given deviation as they pass through the sample.   Thus, dark-field images essentially reflect the composition distribution, and the topography of the solid-liquid interface.  An example of a purely topographic contrast is provided by Figure \ref{dilue}, which shows a dark-field image of  a single-phased solidification front. In this experiment, $\theta_o$ was adjusted for the microscope to capture the light emitted by an  interface normal to $z$.  Non-deformed regions of the front appear bright, and distorted regions  (grooves associated with grainboundaries) dark. Satisfactorily, the image width of large-angle grainboundary grooves is comparable to the known physical width ($\approx 12 \mu \rm{m}$) of such grooves in \al~\cite{bottin02}.   The contrast associated with low-angle  grainboundaries is weak in accordance with the small amplitude of the corresponding grooves.

In the case of eutectic patterns, both the differences in refractive index between the different phases, and the strong local deformation of the interface contribute to the image contrast. Let us consider specifically quasi-stationary fibrous eutectic patterns (Fig. \ref{schemaImage}b). We first note that, when a pattern is stationary, the solid-solid interfaces left behind in the solid are parallel to $z$ so that refraction at these interfaces does not affect the angle between the emergent light rays and $z$. Consequently, the solid behind the interface appears totally dark  in spite of multiple refraction by solid-solid interfaces. Second we remark that, in the type of dark-field images illustrated in Fig. \ref{rods}, each DC fiber gives rise to a diffuse light patch ("bright spot"), which does certainly not coincide with the interface between the fiber and the liquid ("fiber tip"). Let us discuss the nature of these bright spots in some detail.

\begin{figure}[hbtp]
 \includegraphics[width=3.5cm]{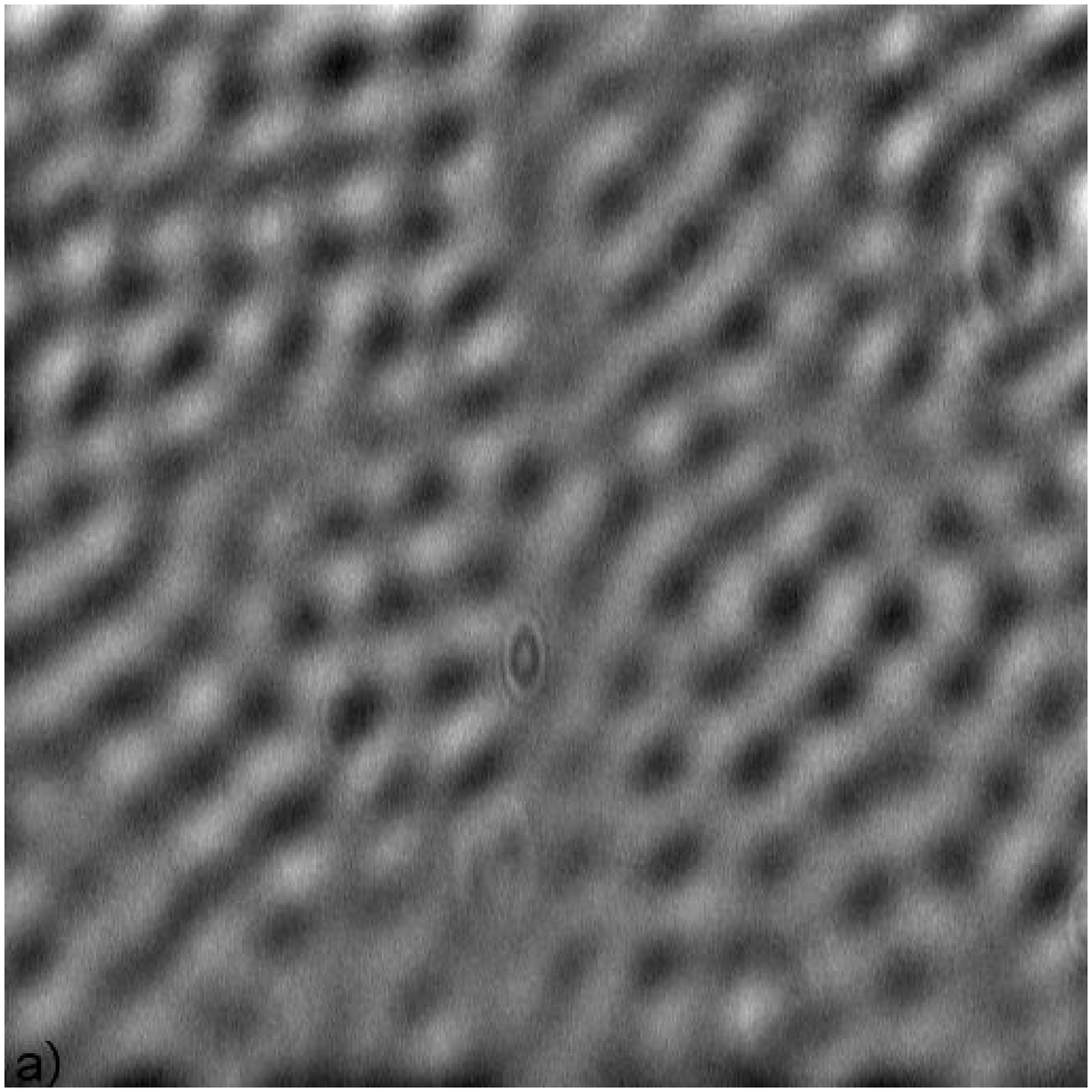}
\includegraphics[width=3.5cm]{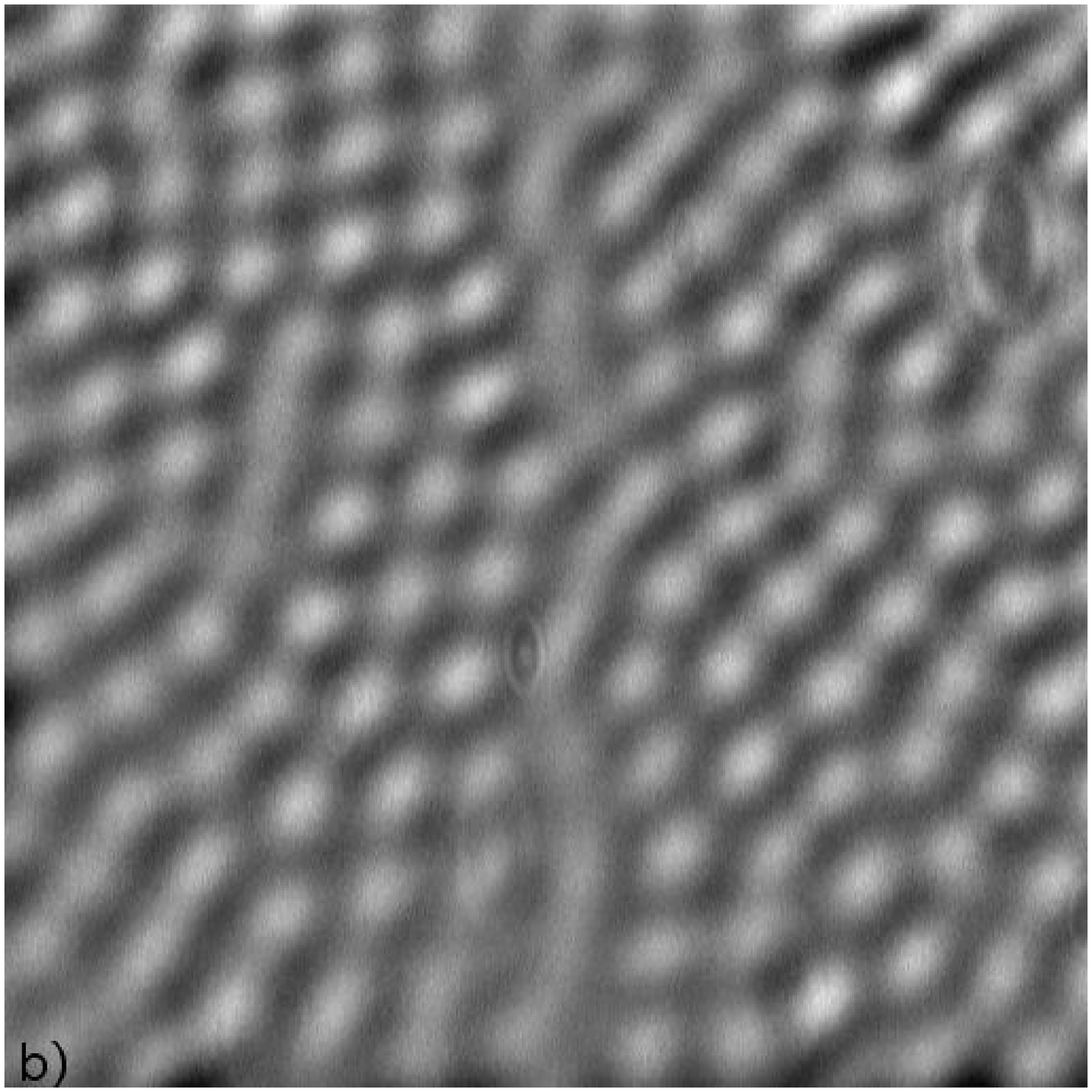}
\caption{Fibrous pattern in a 400-$\mu \rm{m}$ thick eutectic SCN-DC~sample. $V= 0.07 \mu \rm{m.s^{-1}}$. Rescaled images. $\theta_t= 40$\degres. Horizontal dimension: $160 \mu \rm{m}$. a) $\theta_o=38$\degres. b) $\theta_o=41$\degres. The SCN matrix appears bright in a) and dark in b), and vice-versa for the DC fibers.}
\label{InvContraste}
\end{figure}

\begin{figure}
\includegraphics[width=5cm]{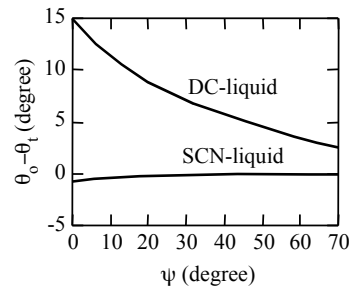}
\caption{Calculated deviation of a light beam intersecting a SCN-, or a DC-liquid interface as a function of the slope angle of the interface for $\theta_t= 40$\degres. } 
\label{deltaTETA}    
\end{figure}

\begin{figure}[htbp]
\includegraphics[width=2.5cm]{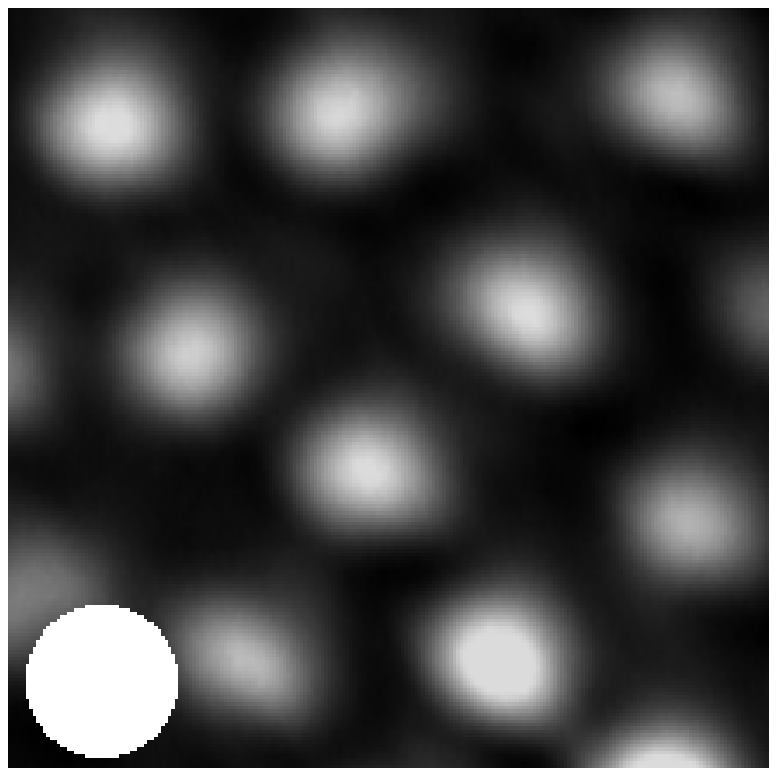} 
\includegraphics[width=2.5cm]{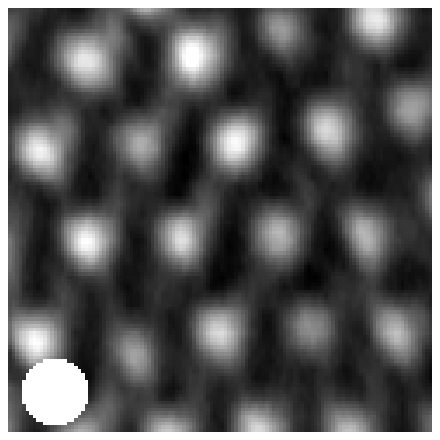} \\
\includegraphics[width=2.5cm]{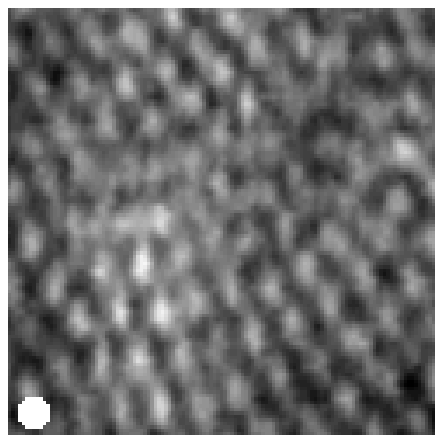}
\includegraphics[width=2.5cm]{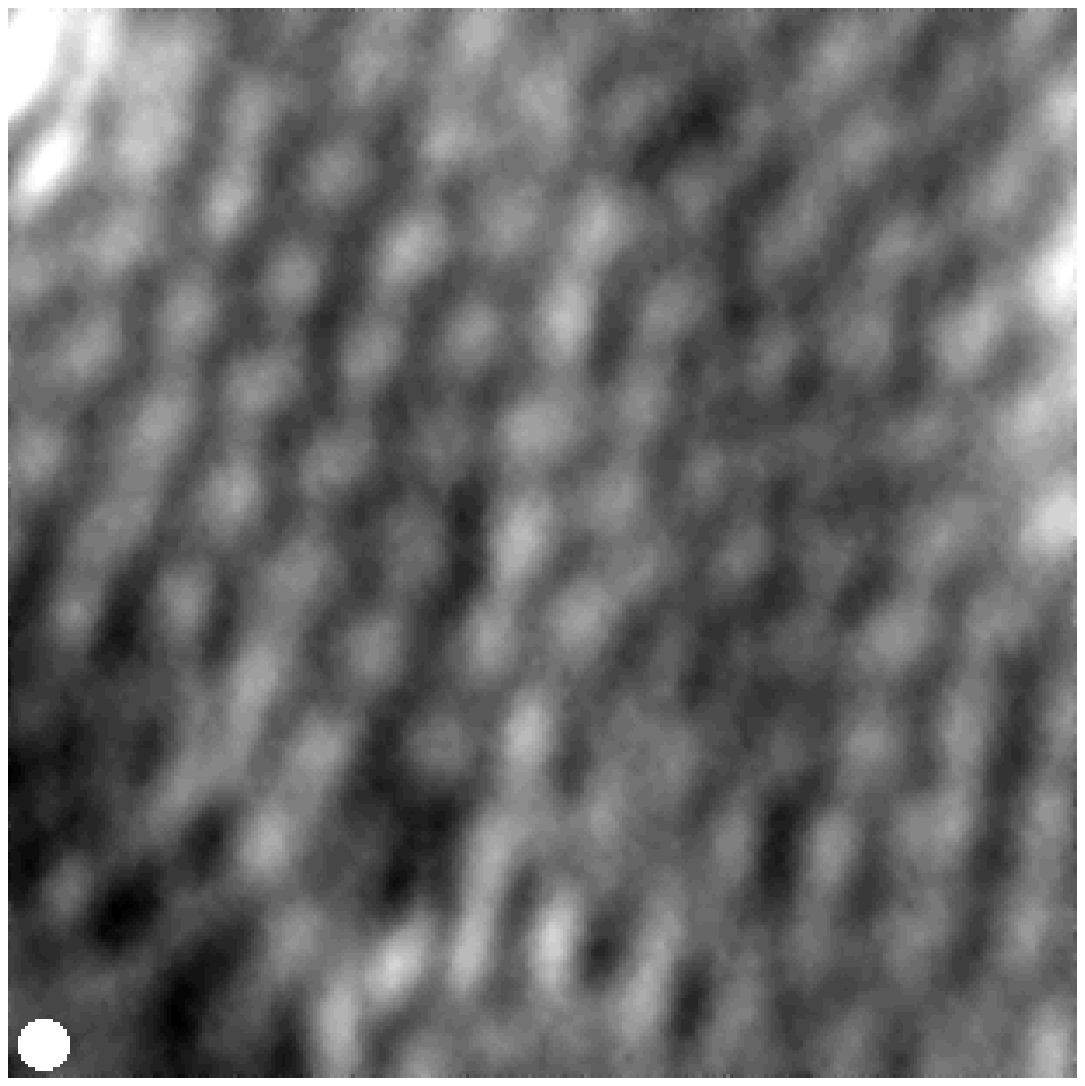}
\caption{Enlarged views of SCN-DC fibrous eutectic patterns at different growth velocities $V$. Rescaled images. White disks: theoretical average cross-section of a DC fiber. The average spacing $\lambda$ decreases as $V$ increases, in accordance with the well-known $\lambda \propto V^{-1/2}$ scaling law of eutectic growth \cite{HuntJacks66}. Left to right, top to bottom: a) $V= 0.018~\mu \rm{ms^{-1}}$; $\lambda \approx 22~\mu \rm{m}$; b)   $V= 0.053 \mu \rm{ms^{-1}}$; $\lambda \approx 13.5~\mu \rm{m}$;  c)  $V=0.124 \mu \rm{ms^{-1}}$; $\lambda  \approx 7.5~\mu \rm{m}$; d) $V=0.177 \mu \rm{ms^{-1}}$; $\lambda  \approx 6.1~\mu \rm{m}$. Horizontal dimension of each micrograph: 70 \mic. }
\label{quality}
\end{figure}

In the case of SCN-DC, an inversion of the contrast between fibers and matrix is observed as a function of $\theta_o$ at fixed $\theta_t$ (Figure  \ref{InvContraste}). The maximum of brightness is at about $\theta_o=\theta_t-2\rm{^o}$ for the SCN matrix, and $\theta_o=\theta_t+1\rm{^o}$ for the DC fibers. Figure \ref{deltaTETA} shows the calculated value of  $\theta_o-\theta_t$  as a function of $\psi$  at fixed $\theta_t$ for the two solid  phases. The calculated values are in reasonable agreement with the experimental ones  for  $\psi \approx 60\rm{^o}$, thus for a solid-liquid interface roughly normal to the light rays. On the other hand, we compared directly the area of the bright spots with the theoretical value  $A_{DC}$ of the fiber cross-section. In a regular hexagonal pattern, $A_{DC}= \eta \sqrt{3}  \lambda^{2}/2$, where $\eta \approx 0.23$ is the equilibrium volume fraction of DC in the solid, and $\lambda$  is the spacing. Figure \ref{quality} shows enlarged views of a series of SCN-DC fibrous patterns at average spacings ranging from 22 to 6 \mic, on which white disks of areas $A_{DC}(\lambda)$ have been superposed. It can be seen that the area of the bright spots, although it is ill defined, is clearly smaller than $A_{DC}$, at least when $\lambda$ is sufficiently large. We conclude that the bright spots are  caustics, which are generated by the rounded tips of the fibers, and are concentrated around a direction making a small angle with the direction of the incident beam.

Fig. \ref{quality} also shows that the ratio of the diameter of the bright spots to the actual fiber diameter  increases as the fiber diameter decreases. This is due to the diffraction-induced broadening of the bright spots. The value of the  fiber diameter, at which the ratio is unity, can be considered as the empirical resolution power of the instrument. From Fig. \ref{quality}, we estimate this resolution power to be  of about  3.5\mic, which is somewhat larger than the diffraction-limited resolution of the LD microscope (2.2~\mic). Previous studies demonstrated a similar empirical resolution power for \al~lamellar eutectic patterns \cite{zigzag,philmag}.

\section{Thermal-bias correction} \label{thermalbias}

In a perfect directional-solidification experiment, the isotherms would be planar and perpendicular to the solidification axis. In any real experiment, the isotherms are both curved and, on average, tilted with respect to the solidification axis (thermal bias). The fact that the dynamics of solidification patterns is very sensitive to these instrumental imperfections has not always received sufficient attention. A thermal bias progressively drives the structure along the direction of the bias, while a curvature causes a progressive blow-up (when the isotherms are bulging outward from the solid, which is our case) of the structure over time.  This will be the object of a detailed report in the near future \cite{Perrut}.  Here we content ourselves with stressing the importance of these effects \cite{remark}.

We determined the curvature of the isotherms in our setup by following the dynamics of fibrous eutectic patterns over long periods of time. We found that the curvature was negligible in the direction $x$, but not in the direction $y$.  The radius of curvature in the direction $y$ ranged from 2 to 6~mm in 400-\mic~thick samples, depending on the experiment, which corresponds to a difference in $z$  between two points located at $y=0$ and $y=w/2$, respectively, ranging from 3 to 10\mic. There is apparently no way to correct this curvature, which mostly arises from the discontinuities in heat conductivity at the solid-liquid, solid-wall and liquid-wall interfaces.

\begin{figure}[htbp]
\includegraphics[width=7cm]{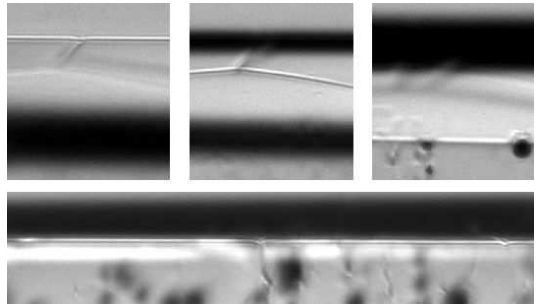}
\caption{Side view of a 400-$\mu \rm{m}$ thick sample of a dilute \al~alloy at rest. Upper panel: A focal series before thermal-bias correction. Horizontal dimension of a micrograph: $130 \mu \rm{m}$. Left to right: the focus changes from the base, to the cover glass plate. In this example, $\Delta z=83 \mu \rm{m}$ and $\psi_b=12\rm{^o}$.  Lower panel:  Same interface after  thermal-bias correction. The focus is on the cover glass plate. Horizontal dimension: $420 \mu \rm{m}$.  }
\label{biais}
\end{figure}

We determine the thermal bias at the onset of an experiment by observing the solid-liquid interface at rest ($V=0$) in side view, and correct it by changing the temperature of the hot cover block in the DS bench. The details of the procedure are as follows. Previous to a solidification run,  a sample undergoes a partial directional melting, after which it is held at rest for equilibration. When the annealing is completed, the concentration $C_L$ in the liquid is homogeneous, and the shape of the solid-liquid interface is that of the isotherm at the liquidus temperature $T_L$ of the alloy. (In the case of a eutectic alloy, $T_L$ is the  liquidus temperature for the primary eutectic phase). At this stage, we generally observe a noticeable transverse thermal bias, which we measure using focal-series reconstructions (Fig. \ref{biais}).  The thermal bias is given by  $\psi_b=\arctan (\Delta z/w)$, where $\Delta z$ is the difference in height between the intersections of the interface with the two glass plates. The residual thermal bias after correction does not exceed $2\rm{^o}$. The validity of the procedure can be crosschecked in oblique view, by comparing the measured value of the contraction factor to the theoretical value (eq.  \ref{fperp}).

\section{Conclusion} \label{Conclusion}

We have tested, and validated, a new  method of observation of transparent eutectic growth fronts in bulk samples. The method makes use of three basic ingredients: a microscope with a very long working distance and a narrow aperture, an oblique direction of observation, and a dark-field contrast relying on specific features of eutectic patterns, namely, the difference in optical index between the different solid phases, and the strong local curvature of the interface. With a prototype instrument based on this method, we obtained satisfactory images of eutectic patterns of spacing no larger than 3.5\mic~in 400-\mic~thick samples. Much larger sample thicknesses could probably be utilized with more sophisticated instruments based on the same method. More importantly, the principle of a long-distance oblique observation (without the dark-field specification) could be applied to other objects than eutectic growth patterns, for instance, grain boundaries intersecting the single-phased growth fronts, as shown above, and thus become a standard method in solidification science.

\section{Acknowledgments}
We thank  V.T. Witusiewicz, L. Sturz, and S. Rex from ACCESS (Aachen, Germany) for kindly providing us with purified  Succinonitrile-(d)Camphor  alloy.  This work was supported by the Centre National d'Etudes Spatiales, France.


\begin{thebibliography}{99}
\bibitem{remark} As a method of observation of eutectic growth fronts in bulk samples, some authors purposedly  tilted the isotherms to a large angle while maintaining the microscope in a side-view position. This method is interesting, but it must be kept in mind that it generates a  rapid transverse drift of the pattern, and is thus far from being equivalent to method presented here.
\end{thebibliography}
\end{document}